# Shesop Healthcare: Android Application to Monitor Heart Rate Variance, Display Influenza and Stress Condition using Polar H7

Andrien Ivander Wijaya[*1], Ary S. Prihatmanto[*2], Rifki Wijaya[*3]

[*]School of Electrical Engineering and Informatics
Institut Teknologi Bandung, Jalan Ganesha 10, Bandung 40132, Indonesia
[1]`andrien@outlook.com`
[2]`asetijadi@lskk.ee.itb.ac.id`
[3]`rifkiwijaya@gmail.com`

*Abstract*—Shesop is an integrated system to make human lives more easily and to help people in terms of healthcare. Stress and influenza classification is a part of Shesop's application for a healthcare devices such as smartwatch, polar and fitbit. The main objective of this paper is to create a proper application to implement the stress and influenza classification. The application use Android studio, XML and Java. Also, while creating this application, all design and program is considered to be available for future updates. The application needs an android smartphone with Bluetooth Low Energy technology (bluetooth v4.0 or above). SheSop application will accommodate data entry, device picker, data gathering process, result and saving the result. In the end, we could use the polar H7 and this application to get a real-time heart rate, Heart rate variability and diagnose our stress and influenza condition.

## I. Introduction

Healthcare is a core of human's life. Being healthy is one of the main objective of life ever since BC. According to Oxford's English dictionary, health is the state of being free from illness or injury and that's why for more than a thousand years, human always constantly seek the cure for all disease that spread among the world. Nowadays, everything based on mobility and application based. Smart accommodation, food, government, economy, education, payment and even smart health. More than 4 billion cell phones user from 7.4 billion population is a sign that more than half human is a cellphone user and it could reach more than 80% is the metropolis or huge city.

Recently, almost all cellphone, services and media provider recently launch their smart devices with their own application such as Apple watch, Samsung gear, LG Urbane, ASUS zenwatch, Motorola moto 360, Fitbit smartwatch and polar H7. Those devices always includes a heart rate sensor for the healthcare. Usually, the application will get the heart rate and step taken data from the user and show the result in numbers. The result could be HR graph, step remaining and some application determine whether you are sleeping. But, heart rate data can be used for determine more human's behavior and that's the reason why Shesop application for influenza and stress is created.

## II. Background

### A. Heart Rate Attributes

Heart rate is a pulse beats (Ventricle Contraction) per certain time. Usually, we used 60 seconds to measure heart rate. Heart rate is a dynamic variable which changes over time. HR use bpm which can be measured by count the total beats in 1 minute period. Heart beats can be found in the chest, arm, neck, feet and many other places. We can extract IBI from RR interval (often called normal to normal NN). RR interval is the time difference between two R peaks in ECG signals. Ectopic beats is the abnormal heart beats that lead a sudden spike to ECG graph. If we ignore ectopic beats, we can assume that RRI equals to IBI.

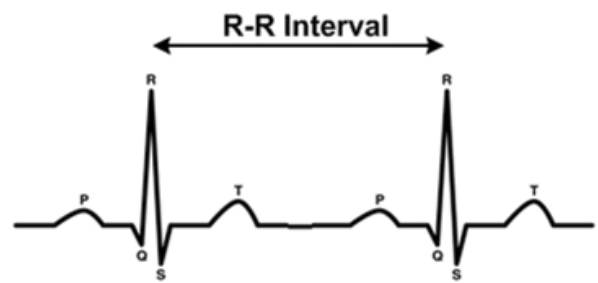

Image 1: ECG signal and PQRST

There are more heart rate variance such as mean, standard deviation, Poincare attribute and entropy.

*B. Devices and Software for Development*

There are some devices which measure heart rate from chest, arm and ear. Technically, chest devices have some advantage than arm devices since it's directly target the heart beats rather than target the veins contraction since veins contraction usually have 1-2 seconds delay from the heart beats. Also, chest strap have no missing data compared with arm wrist. But the arm devices also have some advantage such as easier to use, more comfortable and good mobility while chest devices is harder to use and less comfortable for certain people.

In this paper, we will use polar H7 devices because it's more accurate to do research using chest devices and Polar devices is an open source. Below is the polar H7 device:

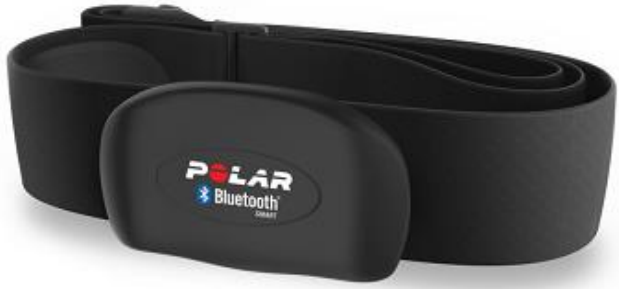

Image 2: Polar H7 device

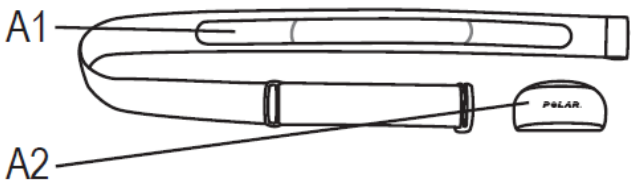

Image 3: Polar H7 sketch

The strap includes ECG electrodes (Image 2 point A1) sensor that detect the heart rate by measuring the electrical activity of the heart to deliver continuous and resting heart rate data and then send the detection data to the module (Image 2 point A2). The module includes bluetooth low energy that send the data to our device such as phone/ tablet via application.

For the software engine, there is two powerful engine for android creation: Android Studio and Unity with to consideration:

| Factors | Android studio | Unity |
| --- | --- | --- |
| UI | Easier with inheritance templates of Activity, stiff, use XML file for UI | Create from zero. Customable rect UI, more flexible UI asset, Use Scene for UI. |
| Compatibility | Any SDK | Any SDK |
| Size | Much lighter and smaller | Big and more heavy |
| Language | Java and XML | C# or JavaScript |
| BLE Plugins | Available in java language | Must create plugin for C# |
| Ease of use | Create own frame update, internal script writing, 2D only application | Monobehavior script inheritance support frame update, use visual studio, 3D |
| Battery usage | Application cost battery usage | More heavy for battery |

Table 1: Android Studio VS Unity

Since the application need more UI based with button and text, need BLE plugin, prefer for less size and need lower battery cost rather than a complete feature such as animation and particle, Shesop application will created via android studio.

*C. Interaction Desain and Good Data*

According to Sharp, Rogers and Preece (2011), Interaction design is designing interactive products to support the way people communicate and interact in their everyday and working lives. Application design should give best user experience (UX). UX is something that user feels, could be immersive, happy, sad, smooth or even frustrating. Application also must have good user interface (UI) such as nice button position, feedback response, simple color mixture and readable font. Interaction design should fulfill:

| | |
| --- | --- |
| Effective to use | Effective is about doing the right things. Doing wrong things for a long time could be very ineffective. |
| Efficient to use | Efficient is about doing things right, fast and optimized, according to the SOP or guide, no matter its right or wrong. |
| Safe to use | Save, good security, have rating, have a spoiler alert to shocking or explicit contents |
| Have good utility | Everything must have a utility or at least: we have a future plan for these thing. |
| Easy to learn | Easy to learn even without a proper tutorial. |
| Easy to remember how to use | After several usage, user can remember and use the function properly |
| Have Visibility | Is this a button? What should I do when I want to exit? Everything must be visible and clear (except if you want to make a surprise ) |
| Have Feedback | Have feedback on any action: pressing button, saving, time out, etc. Includes sound, highlighting, animation, vibration and combinations of these. |
| Have Constraints | Restricting the possible actions that can be performed, helps prevent user |
| | from selecting incorrect options. Physical objects can be designed to constrain things. |
| Logical and not ambiguous | Design and language must not be ambiguous and logical. For example, the button uses arrow icon for both back and rewind but only different color. It will lead to an ambiguous application and could be fatal. |
| Consistent | Consistent UI, consistent information. When we break the consistency, it could increases learning burden on user, making them more prone to errors. |
| Good Affordance | Give an attribute of an object that allows people to know how to use it. For example, horizontal scroll will lead to left and right moves and vertical scroll will lead to up and down moves. |

Table 2: Interaction Design

Good design also need to be supported by good data. Data by itself is an indefinite or generic term that needs descriptors to modify and make clear its meaning. There's so many example of data such as: employee data, salary data, purchase data, data on SD card or even data about our favorite video on YouTube. But what makes it good? British scientist William Thompson, Lord Kelvin is quoted as having said: "Until you can measure something and express it in numbers, you have only the beginning of understanding". So, a good data must have a good background and specifications such as: What did we measure? Referenced against what? The dimensions? Consistent enough? Sample is valid enough?

Good data also often linked to relevant, current, and trustworthy business information while bad data often linked with out-of-date, obsolete and low-value information. Good data should consider every simple but significant external factor such as weather, human's mood, temperature, etc. Also, we need to think how the data changed over time and most importantly: the data objective.

## III. APPLICATION DESIGN

### A. Application Initial Design

The application design is:

| | |
|---|---|
| Goal | Input user's condition<br>Pick devices<br>Get heart rate<br>Get HRV result<br>Give classification<br>Upload files to dropbox |
| Target Age Rating | [Teens] Designed to be used by teens and above. Could be used by kids under 10 years with parent's guidance |
| UI | Simple black and white<br>Background gradient |
| Compatibility | Android device<br>API 18 (4.3 Jellybean)<br>Bluetooth 4.0 |
| APK Size | Max 10 MB |
| Class | 9 main class:<br>splash screen class,<br>scan device class,<br>main menu class,<br>device recording and display class,<br>math class,<br>global variable class,<br>bluetooth class,<br>dropbox class and<br>SVM class |
| Layout | 4 main layout XML file:<br>splash screen XML,<br>main menu XML,<br>device pick XML and<br>recording-result XML |
| Android Testing Passed | UX-B1 (Standard design)<br>UX-N1 (Navigation)<br>UX-N2 (Navigation)<br>UX-N3 (Navigation)<br>UX-S1 (Notification)<br>UX-S2 (Notification)<br>FN-P1 (Permission)<br>FN-P2 (Permission)<br>FN-L1 (Install Location)<br>FN-A1 (Audio)<br>FN-A2 (Audio)<br>FN-A3 (Audio)<br>FN-A4 (Audio)<br>FN-S1 (User/app state)<br>FN-S2 (User/app state)<br>FN-U1 (UI and Graphics)<br>FN-U2 (UI and Graphics)<br>FN-U3 (UI and Graphics)<br>PS-S1 (Stability)<br>PS-P1 (Performance)<br>PS-M1 (Media)<br>PS-V1 (Visual quality) |

Table 3: Design and Testing Application

## B. Application Flow Chart

Image 4: Shesop application flowchart

## IV. APPLICATION RESULT AND FEATURE

The application result is shown below:

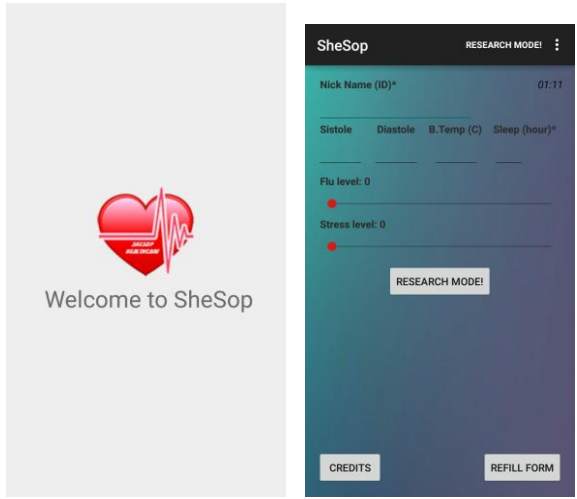

Image 5: Splash screen and data input menu

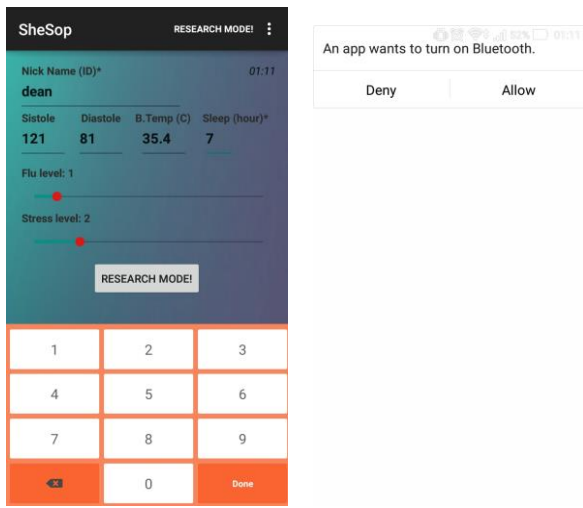

Image 6: input the data and turn on bluetooth notification if it's off

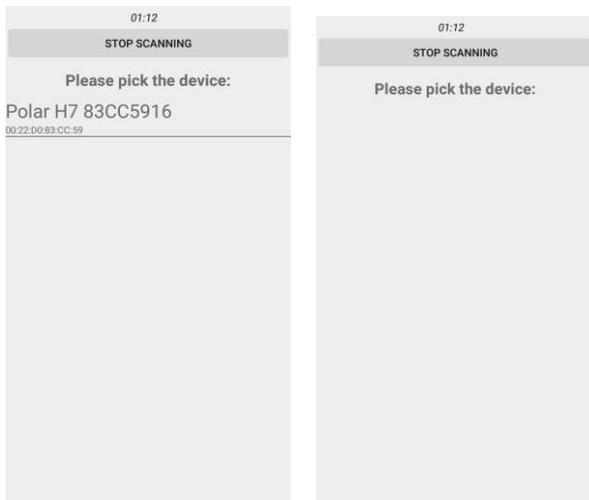

Image 7: pick device menu with polar detected and not detected

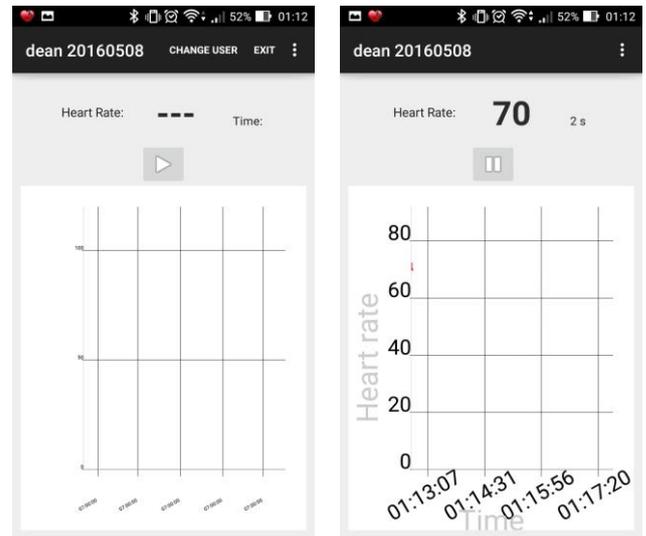

Image 8: Recording menu and start recording

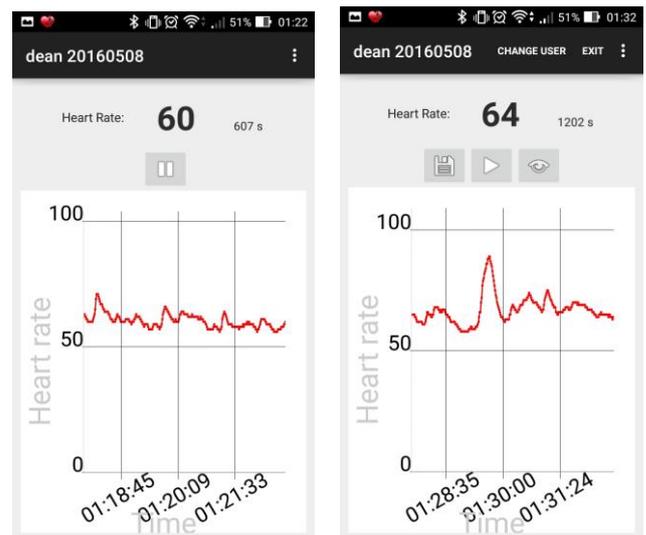

Image 9: 607 and 1202 seconds of recording

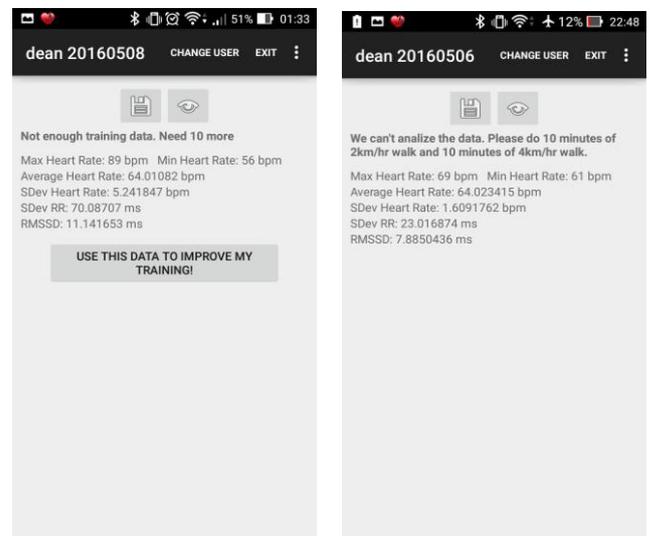

Image 10: Result if not enough data or training duration is too short

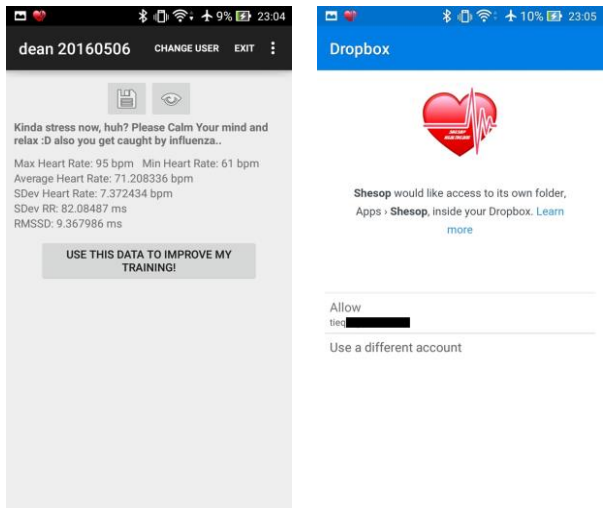

Image 11: Result and dropbox upload

The best application is the application that need to be improved periodically. While finishing this paper, the application was on the 7th revision.

## V. Conclusion

Heart rate can be measured from certain place such as chest, arms and neck. Heart rate can be related to the body's condition and mentality. Because HR can be changed very dynamically, data should be taken in a stable condition. In this application, data will be taken using polar H7.

Heart rate variance can be used to analyze the human body. HRV includes Time domain, frequency domain, poincare, non-linear and time frequency. The result will be shown in the view result mode after the training session is done.

A good application will consider the user, design, future update and clean code. Proper Shesop application has been made and passed the test for android application. Using the classification that has been made before, this application will be used to help people identify their condition and hopefully in future can classify more body condition.

## References

[1] Andrien Ivander Wijaya, Ary S Prihatmanto, Rifki Wijaya. 2016. Shesop Healthcare: Stress and influenza Classification Using Support Vector Machine Kernel.

[2] European Heart Journal. 1996. Heart rate variability: Standards of measurement, physiological interpretation, and clinical use.

[3] Gary G Berntson. 1997. Heart rate variability: origins, methods and interpretive caveats.

[4] Jenifer Tidwell. 2010. Designing Interfaces. O'Reilly Media: Canada.

[5] Juan F. Ramirez-Villegas, Eric Lam-Espinosa, David F. Ramirez-Moreno. Heart Rate Variability Dynamics for the Prognosis of Cardiovascular Risk

[6] J.T. Ramshur. 2010. HRVAS: Heart Rate Variability Analysis Software University of Memphis.

[7] J.T. Ramshur. 2010. "Design, Evaluation, and Application of Heart Rate Variability Analysis Software (HRVAS)". University of Memphis.

[8] Markad V. Kamath, Mari A. Watanabe, Adrian R.M. Upton. 2012. Heart Rate Variability (HRV) Signal Analysis: Clinical applications.

[9] Melissa Webster. 2015. Big Data, Bad Data, Good Data: The Link Between Information Governance and Big Data Outcomes. IBM.

[10] Paolo Melillo, Marcello Bracale and Leandro Pecchia. 2011. Nonlinear Heart Rate Variability features for real-life stress detection. Case study: students under stress due to university examination.

[11] Yvonne Rogers, Helen Sharp, Jenny Preece. 2011. Interaction Design: Beyond Human - Computer Interaction.

[12] http://www.android.com/ , 2016

[13] http://www.fitbit.com/ , 2016

[14] http://www.marcoaltini.com , 2016

[15] http://www.mathworks.com/ , 2016

[16] http://www.heart.org , 2016

[17] http://www.polarusa.com/ , 2016